\documentclass{article}

\usepackage{graphicx}      
\usepackage{cite}        

\usepackage{amsmath}
\usepackage{multirow}
\usepackage{caption,subcaption}
\usepackage{enumerate}
\usepackage{xcolor}
\usepackage{amssymb}
\usepackage{accents}
\newcommand{\ubar}[1]{\underaccent{\bar}{#1}}

\usepackage{amsthm}
\usepackage{authblk}
\usepackage{hyperref}

\newcommand{\mycomment}[1]{}

\newtheorem{proposition}{Proposition}
\newtheorem{problem}{Problem}

\newtheorem*{pf}{Proof}
\newtheorem{thm}{Theorem}

\begin{document}

\title{Hydraulic Parameter Estimation in District Heating Networks} 
\date{}
\author[1]{Felix Agner\thanks{Corresponding author: \href{mailto:felix.agner@control.lth.se}{felix.agner@control.lth.se}}\thanks{This work is funded by the European Research Council (ERC) under the European Union's Horizon 2020 research and innovation program under grant agreement No 834142 (ScalableControl).}}
\author[2]{Pauline Kergus}
\author[1]{Richard Pates}
\author[1]{Anders Rantzer}

\affil[1]{Automatic Control Department, Lund University, Sweden}
\affil[2]{LAPLACE, Université de Toulouse, CNRS, INPT, UPS, Toulouse, France}

\maketitle

\begin{abstract}                
Using hydraulic models in control of district heating networks can increase pumping efficiency and reduce sensitivity to hydraulic bottlenecks. These models are usually white-box, as they are obtained based on full knowledge of the district heating network and its components. This type of model is time-consuming to obtain, and might differ from the actual behavior of the system. In this paper, a method is proposed to obtain a grey-box hydraulic model for tree-shaped district heating systems: hydraulic parameters are estimated based on pressure measurements in only two locations. While previous works only estimate parameters related to pressure losses in pipes, this work also includes customers valves in the grey-box model structure, an important inclusion for control-oriented applications. Finally, a numerical example illustrates the proposed method on a small district heating network, showing its ability to obtain an accurate model on the basis of noisy measurements.
\end{abstract}

\newpage


\section{Introduction}

District heating networks already play an important role in the energy mix of some countries, such as Sweden and Denmark, see \cite{thebible}. As mentioned in \cite{lund2018status4th}, district heating (and cooling) has an important role to play in the transition to future sustainable energy solutions. As underlined in \cite{vandermeulen_controlling_2018}, it is required to tackle several control challenges to utilize the full potential of district heating in these future systems. In particular, \cite{vandermeulen_controlling_2018} recalls that enabling low supply temperatures and ensuring a fair distribution of heat (and cold) is important in the transition to more efficient district heating networks. Both aspects are linked to the regulation of flow rates. Indeed, lowering the supply temperature can lead to higher flow rates, which may create bottlenecks \cite{brange_decision-making_2019}. Bottlenecks are events of insufficient differential pressure between supply and return pipes in some parts of the network, caused by significant pressure losses. 

In this context, regulating the flow in district heating networks is an important control problem, not only in future generations of district heating but already in existing networks, see \cite{thebible}.
\mycomment{Traditionally, this control relies on valves and pumps. The valves, located at each customer substation, are controlled locally to get the flow needed to meet their demand. The pumps are centrally regulated so that the differential pressure of the network is sufficiently high for the valves to perform their task.} In this flow control perspective, considering a hydraulic model of district heating networks can be beneficial. \cite{hydraulic_optim_meshed} showed that using such a model can reduce the necessary pumping power. \cite{AGNER2022100067} showed that hydraulic models can be used to reduce the effect of bottlenecks in times of peak demand.

Hydraulic modelling is usually white-box, see \cite{thebible}. Pressure losses in pipes are described using lengths, diameters and friction coefficients. Valves are modelled with characteristics provided by manufacturers. Such white-box models require gathering a lot of information, which may be complicated in practice. In addition, there is no guarantee that the model will fit the actual behavior of the district heating network. This paper aims at proposing a method to estimate a grey-box hydraulic model from measured data. Such data-driven modelling allows not only to save time in the modelling process, but also keeps the model accurate with regards to measurements. 

Estimation of hydraulic model parameters has been investigated for water distribution systems, as summarized in \cite{quovadisWDS}. However, only a few works have been published in the specific area of district heating systems, for which the models will differ in structure.
\cite{NetworkParameterEstimationForDHSystem} estimates a hydraulic model based on all nodal pressure head values in the network. \cite{WANG201883} and \cite{DH_pipe_hydraulic_estimation} conducted similar studies, using measurements of pressures at all customer substations. Loop structured graphs are treated in \cite{DH_pipe_hydraulic_estimation} by working with several overlapping tree-structured graphs. These works have shown that hydraulic parameter estimation from data is possible in theory. However, their identification process only accounts for pipes. Customer valves are neglected. Each such valve influences the pressure in the whole network and they are therefore important to consider for flow control purposes. This paper extends the work of \cite{DH_pipe_hydraulic_estimation} and \cite{WANG201883} by introducing a parameterized valve model in the grey-box identification process. In addition, the assumption that nodal heads are measured at every substation is alleviated. This raises the question of whether all parameters in the network can still be uniquely identified.

The contributions of this work are as follows.
\begin{enumerate}
    \item The non-linear hydraulic relations of a tree-structured district heating network are reduced to a set of equations which are linear with regards to the parameters under consideration.
    \item The conditions for these equations to have a unique solution are provided, analyzed, and shown likely to hold in practice. This allows parameter estimation to be performed with a simple least square solution.
    \item A numerical example demonstrates the feasibility of the method, based on noisy measurements.
\end{enumerate}

Section \ref{sec:network structure} introduces the notation used to describe the considered networks, along with constraints related to flows and pressure in the network. The linear set of equations are presented in section \ref{sec:linear constraints and estimation method}. The conditions for a unique solution to the equations is analyzed in section  \ref{sec:uniqueness of solution}. Finally the numerical example is presented in section \ref{sec:numerical example}.


\section{Network Constraints}\label{sec:network structure}
This section introduces notation for describing the considered networks, followed by models and constraints related to flows and pressures in the network.

\subsection{Network Description}
The type of network under consideration consists of two directed sub-graphs, a \textit{supply-network} $\Bar{\mathcal{G}} = (\Bar{\mathcal{E}},\Bar{\mathcal{N}})$ and a \textit{return-network} $\ubar{\mathcal{G}} = (\ubar{\mathcal{E}},\ubar{\mathcal{N}})$. They are symmetric in structure. In addition, there is a set of \textit{boundary edges} $\mathring{\mathcal{E}}$ that are connected from some nodes in $\Bar{\mathcal{N}}$ to their equivalent counterparts in $\ubar{\mathcal{N}}$. Define also two reference nodes $\alpha \in \Bar{\mathcal{N}}$ and $\beta \in \ubar{\mathcal{N}}$, and assume all edges in $\Bar{\mathcal{E}}$ and $\ubar{\mathcal{E}}$ to be directed away from $\alpha$ and $\beta$ respectively. Non-reference nodes which are not connected to a boundary edge are called \textit{internal nodes}, and are assumed to have at least degree three. Denote a path $\mathcal{P}$ to be a set of ordered edges traversing a set of ordered, distinct nodes. 

\begin{figure}
    \centering
    \includegraphics[width = .65\columnwidth]{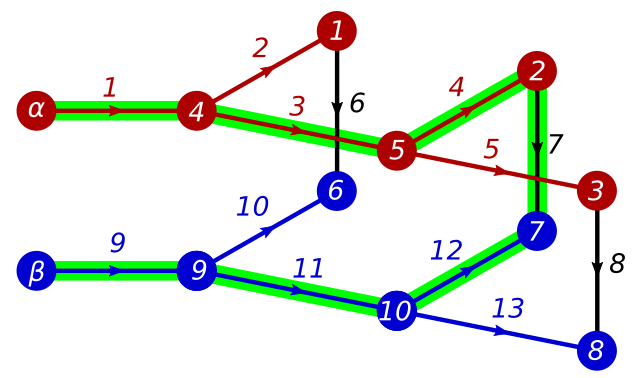}
    \caption{Small example network illustrating the network structure. Red edges and nodes are part of the supply-network, blue edges and nodes are part of the return network, and the black edges are the boundary edges. The green highlight shows a path $\mathcal{P}$.}
    \label{fig:small example network.}
\end{figure}

Figure \ref{fig:small example network.} illustrates such a network. Here $\Bar{\mathcal{N}} = (\alpha,1,2,3,4,5)$, $\ubar{\mathcal{N}} = (\beta,6,7,8,9,10)$, $\Bar{\mathcal{E}} = (1,2,3,4,5)$ $\ubar{\mathcal{E}} = (9,10,11,12,13)$ and $\mathring{\mathcal{E}} = (6,7,8)$. Internal nodes are $4$, $5$, $9$ and $10$. The highlighted path $\mathcal{P}=(1,3,4,7,12,11,9)$ traverses the nodes $(\alpha,4,5,2,7,10,9,\beta)$.

Associate each edge $e$ with a volumetric flow rate $q_e$. Denote the vectors of such flows $\Bar{q}$, $\ubar{q}$ and $\mathring{q}$ for supply, return and boundary edges respectively. Associate with each node $n$ a pressure-value $p_n$. Denote the vectors of such pressures $\Bar{p}$ and $\ubar{p}$ for supply and return nodes respectively. Associate with each edge $e$ a control value $u_e \in \left]0,1\right]$, and let $u_e=1$ for $e\in \Bar{\mathcal{E}} \cup \ubar{\mathcal{E}} $. An edge $e$ also has an unknown resistance parameter $s_e$, which is the parameter we seek to estimate in this paper. Denote $\Bar{s}$, $\ubar{s}$ and $\mathring{s}$ to be the vectors of such parameters related to edges in $\bar{\mathcal{E}}$,  $\ubar{\mathcal{E}}$ and $\mathring{\mathcal{E}}$ respectively. Assume symmetry between characteristics of supply and return edges such that $\bar{s} = \ubar{s}$.

\subsection{Flow Model}
Each node is subject to preservation of mass, i.e. the sum of incoming flow must equal the sum of outgoing flow. Considering first the supply-network, this can be equivalently written as
\begin{equation}
    \begin{bmatrix}
    \mathring{q} \\ 0
    \end{bmatrix} = \Bar{B} \Bar{q}
    \label{eq:flow equality}
\end{equation}
where $\Bar{B} = \left[ \Bar{B}_{i,j} \right] $ is the basic incidence matrix of $\Bar{\mathcal{G}}$, defined as
\begin{equation}
    \Bar{B}_{i,j} = \begin{cases} 
    1, & \text{if edge j leads to node i} \\
    -1,& \text{if edge j leads away from node i} \\
    0,& \text{if edge j is not connected to node i.}
    \end{cases}
\end{equation}
Here the node $\alpha$ is omitted to ensure \eqref{eq:flow equality} is not over-determined. Note that in \eqref{eq:flow equality}, it is assumed without loss of generality that the nodes are ordered such that internal nodes come last. See e.g. the network visible in Figure \ref{fig:small example network.} for which
\begin{equation}
    \Bar{B} = \begin{bmatrix}
    0 & 1 & 0 & 0 & 0 \\
    0 & 0 & 0 & 1 & 0 \\
    0 & 0 & 0 & 0 & 1 \\
    1 & -1 & -1 & 0 & 0 \\
    0 & 0 & 1 & -1 & -1 \\
    \end{bmatrix}.
\end{equation}
Due to the tree-structure of $\Bar{\mathcal{G}}$, $\Bar{B}$ is invertible. This means that given known boundary flows $\mathring{q}$, the supply-flows $\Bar{q}$ can be calculated as
\begin{equation}
\Bar{q} = \Bar{B}^{-1}\begin{bmatrix}
    \mathring{q} \\ 0
    \end{bmatrix}.
    \label{eq:supply flows}
\end{equation}
Due to symmetry between $\Bar{\mathcal{G}}$ and  $\ubar{\mathcal{G}}$, the corresponding basic incidence matrix $\ubar{B}=\Bar{B}$. This in combination with the boundary flows $\mathring{q}$ being directed from $\Bar{\mathcal{G}}$ to $\ubar{\mathcal{G}}$ gives
\begin{equation}
    \ubar{q} = - \bar{q}.
    \label{eq:supple-return-flow}
\end{equation}

\subsection{Differential Pressure Model}
For an edge $e$ with flow $q_e$, directed from node $n$ to node $m$ with pressures $p_n$ and $p_m$, models of the following type are considered.
\begin{equation}
    p_n-p_m = f(q_e,u_e)s_e,
    \label{eq:pressure two nodes}
\end{equation}
where $s_e$ is the unknown resistance parameter to estimate and $f: \mathbb{R} \times \left]0,1\right] \rightarrow{} \mathbb{R}$ is a given continuous function satisfying
\begin{align}
    & f(q,u) = -f(-q,u), \label{eq:f symmetrical}\\
    & f(0,u) = 0, \label{eq:f(0,u)=0}\\
    & q_i > q_j \implies f(q_i,u) > f(q_j,u), \label{eq:f monotone q}\\
    & u_i > u_j \implies f(q,u_i) < f(q,u_j), \label{eq:f monotone u}\\
    & q > 0 \implies \lim_{u \to 0} f(q,u) = \infty \label{eq:f unbounded}.
\end{align}
A specific choice of $f$ will be presented in section \ref{sec:numerical example}, where a numerical example is investigated.

Consider all paths $\mathcal{P}_v$ that lead from $\alpha$ to $\beta$ via a boundary edge $v \in \mathring{\mathcal{E}}$. Denote $n_b$ to be the number of such paths, i.e. the number of boundary edges. The path $\mathcal{P}$ in Figure \ref{fig:small example network.} is an example of such a path. Then
\begin{equation}
    p_\alpha - p_\beta = \sum_{e\in \mathcal{P}_v \cap \Bar{\mathcal{E}}} f(q_e,1)s_e - \sum_{e\in \mathcal{P}_v \cap \ubar{\mathcal{E}}} f(q_e,1)s_e + f(q_v,u_v)s_v \label{eq: path alpha beta}
\end{equation}
for all $v$. This equation is the result of subsequent uses of \eqref{eq:pressure two nodes}. Note that $e\notin \mathring{\mathcal{E}} \implies u_e = 1$. The signs in front of the sums are decided by the direction of edges in the supply (return) network being directed from $\alpha$ ($\beta$).

Since $\Bar{\mathcal{G}}$ and $\ubar{\mathcal{G}}$ are symmetrical, the sets $\mathcal{P}_v \cap \Bar{\mathcal{E}}$ and $\mathcal{P}_v \cap \ubar{\mathcal{E}}$ contain the symmetrically corresponding edges. Using \eqref{eq:supple-return-flow}, \eqref{eq:f symmetrical} and the assumption $\bar{s} = \ubar{s}$,
\begin{equation}
    p_\alpha - p_\beta = \sum_{e\in \mathcal{P}_v \cap \Bar{\mathcal{E}}} 2f(q_e,1)s_e + f(q_v,u_v)s_v. \label{eq: path alpha beta simplified}
\end{equation}
Considering \eqref{eq: path alpha beta simplified} for all $n_b$ paths at once yields
\begin{equation}
    Fs = \mathbf{1}(p_\alpha - p_\beta).
    \label{eq:constraints one load condition}
\end{equation}
Here $s = \left[ \Bar{s}^T , \mathring{s}^T \right]^T$ is a vector of $n_s$ resistances and $F = \left[F_{i,j}\right]$ is an $n_b \times n_s$ matrix defined as
\begin{equation}
    F_{i,j} = \begin{cases} 
    2f(q_j,1), & \text{if }j\in\mathcal{P}_i \cap \Bar{\mathcal{E}} \\
    f(q_j,u_j),& \text{if }j\in\mathcal{P}_i \cap \mathring{\mathcal{E}} \\
    0,& \text{else.}
    \end{cases}
\end{equation}

\section{Linear Constraints for Hydraulic Parameter Estimation} \label{sec:linear constraints and estimation method}
This section presents a set of equations which is linear in the parameters $s$, and presents a parameter estimation method based on these constraints.

\subsection{The Parameter Estimation Problem} \label{sec:parameter estimation problem definition}
For parameter estimation, we will use several samples of steady state measurements. Denote any such steady state of pressures, flows and valve positions in the network a \textit{load condition} $t$. For example $\mathring{q}(t)$ is the vector of boundary flows $\mathring{q}$ at load condition $t$. The parameter estimation problem is therefore:

\begin{problem}\label{thm:parameter estimation problem}
Given load conditions $t = 1\dots T$ with measurements of pressures $p_\alpha(1) \dots p_\alpha(T)$ and $p_\beta(1) \dots p_\beta(T)$, controller settings $u(1),\dots, u(T)$ and boundary flows $\mathring{q}(1), \dots , \mathring{q}(T)$, find an estimate $\hat{s}$ of the resistances $s$, such that $\hat{s} = s$.
\end{problem}

\subsection{Linear Constraints}
Let $F(t)$ be defined by \eqref{eq:constraints one load condition} for one load condition $t$. Note that $F(t)$ can be constructed using only the boundary flows $\mathring{q}(t)$ and the controller settings $u(t)$. Using the measurements defined in Problem \ref{thm:parameter estimation problem}, the equation
\begin{equation}
    \begin{bmatrix}
    F(1) \\ F(2) \\ \vdots \\ F(T)
    \end{bmatrix}s = \begin{bmatrix}
    \mathbf{1}(p_\alpha(1) - p_\beta(1)) \\
    \mathbf{1}(p_\alpha(2) - p_\beta(2)) \\
    \vdots \\
    \mathbf{1}(p_\alpha(T) - p_\beta(T))
    \end{bmatrix}
    \label{eq:linear constraints long}
\end{equation}
can be constructed. Let $\Phi$ be the matrix in the left hand side of \eqref{eq:linear constraints long} and $y$ be the right hand side of \eqref{eq:linear constraints long} such that \eqref{eq:linear constraints long} can be condensed to 
\begin{equation}
    \Phi s = y.
    \label{eq:linear constraints short}
\end{equation}
Note that linear constraints were constructed also in \cite{WANG201883} and \cite{DH_pipe_hydraulic_estimation}. Our extension is twofold: They required the measurement of pressure at each boundary node, while we use only $p_\alpha$ and $p_\beta$. In addition we include modeling of the valve characteristics which is important for control applications. 

\subsection{Estimation Method}
If \eqref{eq:linear constraints short} has a unique solution, a solution $\hat{s}$ to problem \ref{thm:parameter estimation problem} can be obtained as 
\begin{equation}
    \hat{s} = \Phi^\dagger y,
    \label{eq:pseudoinverse}
\end{equation}
where $\dagger$ denotes the the Moore-Penrose pseudoinverse. In a real application, $\Phi$ and $y$ will be subject to measurement noise as well as errors in the model assumption, and in this case \eqref{eq:pseudoinverse} chooses the $\hat{s}$ of minimum Euclidean norm which minimizes the Euclidean distance between observed and estimated values for $y$. While \eqref{eq:pseudoinverse} is the method used in this paper, the linear constraints of \eqref{eq:linear constraints short} can be used in other optimization frameworks. For instance, parameter positivity constraints, other cost functions or other parameter regularization methods could be considered.

\section{Uniqueness of solution}
\label{sec:uniqueness of solution}

Equation \eqref{eq:pseudoinverse} solves problem \ref{thm:parameter estimation problem} if the solution to \eqref{eq:linear constraints short} is unique. Showing that this is likely is the topic of this section.

\subsection{Conditions For Unique Solution}

\begin{thm} \label{thm:conditions for uniqueness}
Consider \eqref{eq:linear constraints short} which is a compact form for \eqref{eq:linear constraints long}, and where each row of \eqref{eq:linear constraints long} is constructed through \eqref{eq: path alpha beta} for a given path $\mathcal{P}_v$ and load condition $t$. Then given load conditions $t=1,\dots T$ and paths $\mathcal{P}_v$, from $\alpha$ to $\beta$ through edge $v$ for all $v\in \mathring{\mathcal{E}}$, the following statements are equivalent.
\begin{enumerate}[i]
    \item Equation \eqref{eq:linear constraints short} has a unique solution 
    \item $\nexists \lambda \in \mathbb{R}^{n_s}$, $\lambda \neq 0$, such that $\forall \mathcal{P}_v$,$\forall t$:
\end{enumerate}
\begin{equation}
    \sum_{e\in \mathcal{P}_v \cap \Bar{\mathcal{E}}} 2f(q_e(t),1)\lambda_e + f(q_v(t),u_v(t))\lambda_v = 0
    \label{eq:condition on dependence.}
\end{equation}

\end{thm}

\begin{pf}
Equation \eqref{eq:linear constraints short} will have a unique solution if and only if $\Phi$ has no linearly dependent columns. Condition ii corresponds exactly to the equation
\begin{equation}
    \Phi \lambda = 0,
\end{equation}
i.e. $\Phi$ has a non-trivial right nullspace and thus linearly dependent columns. Since ii dictates that there is no such $\lambda \neq 0$, $\Phi$ must have no linearly dependent columns. $\hfill \blacksquare$
\end{pf}

Theorem \ref{thm:conditions for uniqueness} is stated in terms of each load condition $t$ used in constructing $\Phi$. However, if \eqref{eq:condition on dependence.} does not hold in general, i.e. for any set of load conditions that could feasibly be sampled, it becomes unlikely to hold for all $t$ if $T$ is large. Thus Given general $q_e$, $q_v$, $\theta_v$, we investigate if
\begin{equation}
    \sum_{e\in \mathcal{P}_v \cap \Bar{\mathcal{E}}} 2f(q_e,1)\lambda_e + f(q_v,u_v)\lambda_v = 0
    \label{eq:general condition on dependence}
\end{equation}
holds for any $\lambda \neq 0$. If not, we can conclude that it's unlikely that \eqref{eq:condition on dependence.} holds for all $t$ unless load conditions were sampled with some form of bias.

\subsection{Independence of Boundary Edge Pressure Drops}
\label{sec:boundary edge pressure drops}
Assume that there is a $\lambda \neq 0$ such that \eqref{eq:general condition on dependence} holds. In addition, assume there is a $v$ such that $\lambda_v \neq 0$. That would imply that
\begin{equation}
    f(q_v,u_v) = -\frac{\sum_{e\in \mathcal{P}_v \cap \Bar{\mathcal{E}}} 2f(q_e,1)\lambda_e}{\lambda_v},
    \label{eq:valve pressure dependent lambda}
\end{equation}
i.e. there is a function mapping flows $q_e$ in edges in $\mathcal{P}_v$ to $f(q_v,u_v)$. As a contradiction, consider the following proposition.

\begin{proposition}
    Consider a given boundary flow $\mathring{q} > 0$, a corresponding $\bar{q}$ that satisfies \eqref{eq:supply flows} and $p_\alpha$, $p_\beta$ such that
    \begin{equation}
        f(q_v,1)s_v \leq p_\alpha - p_\beta - \sum_{e\in \mathcal{P}_v \cap \Bar{\mathcal{E}}} 2f(q_e,1)s_e,
        \label{eq:pressure is sufficient}
    \end{equation}
    for all paths $\mathcal{P}_v$ from $\alpha$ to $\beta$ via a boundary edge $v$. Then there exists control settings $0 < u \leq 1$ such that $\mathring{q}$, $\bar{q}$, $u$, $p_\alpha$ and $p_\beta$ fulfill both flow constraints \eqref{eq:flow equality} and pressure constraints \eqref{eq:pressure two nodes} and is thus feasible. In addition, $f(q_v,u_v)$ fulfills 
    \begin{equation}
        f(q_v,u_v)s_v = p_\alpha - p_\beta - \sum_{e\in \mathcal{P}_v \cap \Bar{\mathcal{E}}} 2f(q_e,1)s_e
        \label{eq:pressure drop over valve}
    \end{equation} for all $v$.
    \label{thm:qb independent}
\end{proposition}

\begin{pf}
Since $\bar{q}$ was constructed from \eqref{eq:supply flows}, $\bar{q}$ and $\mathring{q}$ (and thus also $\ubar{q}$) must fulfill the flow constraints \eqref{eq:flow equality} of the system. In addition, since $u_e=1$ and $q_e$ is given for all edges $e$ in the supply-and-return networks, the nodal pressures which solve \eqref{eq:pressure two nodes} are known. Since $f$ fulfills both \eqref{eq:f monotone u} and \eqref{eq:f unbounded}, \eqref{eq:pressure is sufficient} is the condition for the existence of a feasible $u_v$ such that \eqref{eq:pressure two nodes} holds for each boundary edge. If the condition holds, then \eqref{eq:pressure drop over valve} corresponds to just that $u_v$.$\hfill \blacksquare$
\end{pf}

Proposition \ref{thm:qb independent} shows that $f(q_v,u_v)$ is decided by \eqref{eq:pressure drop over valve}, $f(q_v,u_v)$ is a sum of $p_\alpha-p_\beta$ and a function of the flows along the edges of the path. Since Proposition \ref{thm:qb independent} states that any boundary flow vector $\mathring{q}>0$ and sufficiently large $p_\alpha-p_\beta$  is feasible, $f(q_v,u_v)$ is clearly not only a function of the flows, but also $p_\alpha-p_\beta$. This contradicts \eqref{eq:valve pressure dependent lambda}. Thus if there is a $\lambda$ that fulfills \eqref{eq:general condition on dependence}, then $\lambda_v=0$.

\subsection{Independence of Pressure Drops Along Paths}
\label{sec:path edge pressure drops}
If $\lambda_v = 0$ as previously shown, then returning to \eqref{eq:general condition on dependence} and analyzing one edge $i \in \mathcal{P}_v$ yields
\begin{equation}
    f(q_i,1) = \frac{\sum_{e\in \mathcal{P}_v \cap \Bar{\mathcal{E}},e \neq i} 2f(q_e,1)\lambda_e}{\lambda_i}.
    \label{eq:path pressure dependent lambda}
\end{equation}
I.e. there is a function of the other flows $q_e$ in $\mathcal{P}_v$ yielding $f(q_i,1)$. \eqref{eq:f monotone q} gives that $f$ is monotone in $q_i$. This means that \eqref{eq:path pressure dependent lambda} implies that $q_i$ can be decided as a function of the other flows $q_e$ in the path. To contradict this, consider the following proposition.

\begin{proposition}
Consider $\mathcal{\Tilde{P}} = (e_1,e_2,\dots,e_p)$ to be a path along the supply network from $\alpha$ to a node $\gamma \in \bar{\mathcal{N}}$, where $e_1$ is the first edge of $\mathcal{\Tilde{P}}$. Let $q_{1},q_{2},\dots,q_{p}$ be the flows through these edges. Consider edge $i\in \mathcal{\Tilde{P}}$ with flow $q_i$. Then given flows $q_1, \dots q_{i-1}$ and ${q_i+1}\dots q_p$, i.e. the flows of all previous and subsequent edges along $\mathcal{\Tilde{P}}$, then $q_i$ can have any value in the range $q_{i-1} > q_i > q_{i+1}$. If $i$ is the first (last) edge along the path then it instead holds that $q_i > q_{i+1}$ ($q_i > 0$).
\label{thm:independent path flows}
\end{proposition}

\begin{pf}
Assume edge $i$ leads from some node $n$ to some node $m$ and consider equation \eqref{eq:flow equality}, evaluated in nodes $n$ and $m$ respectively.
\begin{align}
    0 &= q_{i-1} - q_i - q_n\\
    0 &= q_i - q_{i+1} - q_m
\end{align}
The signs are dictated by all edges being directed away from $\alpha$. $q_n$ and $q_m$ are the sum of the flows in all edges directed away from nodes $n$ and $m$ respectively which are not part of the considered path. Due to the assumption that each internal node has at least degree 3, $q_n$ and $q_m$ must exist. As $q_n$ and $q_m$ are sums of some components of the boundary edge flows $\mathring{q}$, where Proposition \ref{thm:qb independent} showed that any $\mathring{q} > 0$ is a viable, any $q_n > 0$, $q_m > 0$ are viable. This, along with a rearrangement gives that any
\begin{align}
    q_{i-1} - q_i & >  0 \label{eq:smaller than previous flow}\\
    q_i - q_{i+1} & >  0\label{eq:greater than subsequent flow}
\end{align}
is viable, and simply correspond to different configurations of boundary flows. If $i$ is the first (last) edge of the path, then only \eqref{eq:smaller than previous flow} (or \eqref{eq:greater than subsequent flow}) applies.$\hfill \blacksquare$
\end{pf}

If \eqref{eq:path pressure dependent lambda} would hold, it implies that $q_i$ as analyzed in Proposition \ref{thm:independent path flows} is a function of the other flows along the path. However, as Proposition \ref{thm:independent path flows} states, there is no such function but rather $q_i$ can take any value within a range. The only conclusion is thus that $\lambda_e=0$ for all $e$. 

Section \ref{sec:boundary edge pressure drops} shows that there is no $\lambda$ where $\lambda_v \neq 0$ for any $v \in \mathring{\mathcal{E}}$ such that \eqref{eq:general condition on dependence} holds in general. This section showed that there is no $\lambda_e \neq 0 $ for any $e \in \bar{\mathcal{E}}$ such that \eqref{eq:general condition on dependence} holds in general. This leads to the conclusion that the only solution to \eqref{eq:general condition on dependence} is $\lambda=0$. If load conditions are not sampled with some bias, it is therefore likely that the conditions of Theorem \ref{thm:conditions for uniqueness} should hold, especially if the number of load conditions $T$ is large.

\section{Numerical Example}\label{sec:numerical example}
A numerical example is now presented using a given network topology. Resistance parameter estimates $\hat{s}$ are computed and analyzed based on both noise-free and noisy measurement data. First a specific choice of $f$ is provided. 

\subsection{District Heating Models}
We consider a district heating model example where all edges $e$ in the supply-and-return networks are pipes. They are described by pressure-relations
\begin{equation}
    f(q_e,1) \cdot s_e = q_e |q_e| \cdot \frac{2 f_{d,e} \rho L}{\pi^2D_e^5},
    \label{eq:darcy weisbach}
\end{equation}

where $f_{d,e}$ is the Darcy friction factor, $\rho$ is the density of the water, $L$ is the length of the pipe and $D_e$ is the diameter of the pipe, all assumed to be constant in time and and along the length of the pipe.

All boundary edges are control valves in customer substations, regulating the flow of hot water these customers receive. Control valves $v$ have different characteristics in their mapping from valve position $u_v$ to flow rate $q_v$ at a given differential pressure. In district heating substations, this characteristic is often linear as stated in \cite{thebible}. We express this as
\begin{equation}
    f(q_v,u_v) \cdot s_v = \frac{q_v|q_v|}{u_v^2} \cdot \frac{1}{k_v^2}
    \label{eq:valve pressure difference}
\end{equation}
for valve $v$ with valve setting $u_v$, where $k_v$ is a linear parameter for the relation between flow and valve position.

\subsection{Example Network}\label{sec:example network}
\begin{figure}[t]
    \centering
    \includegraphics[width = .8\columnwidth]{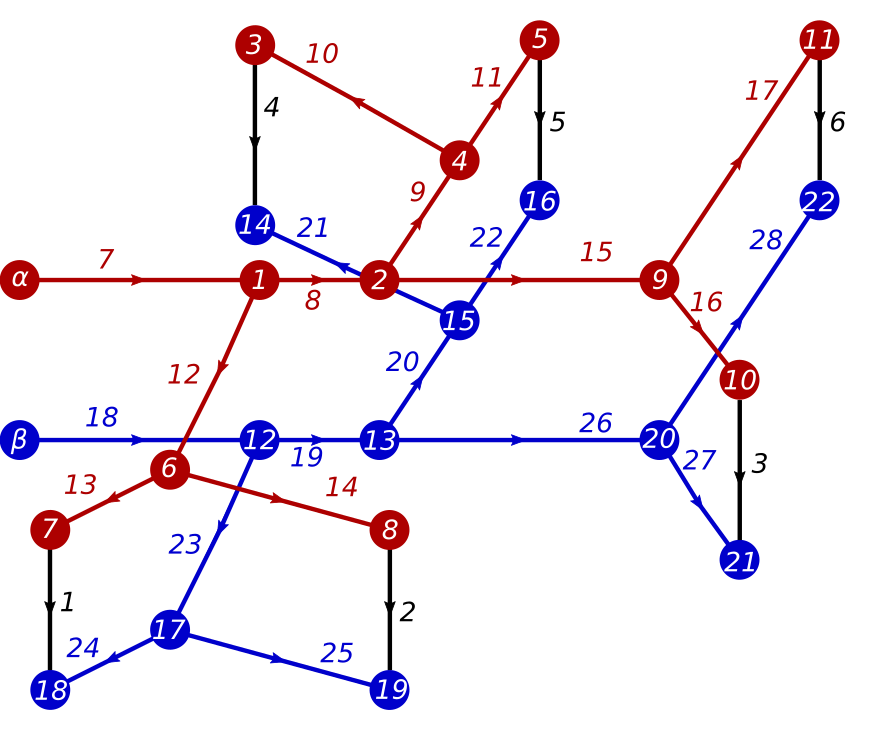}
    \caption{Network used in the numerical example.}
    \label{fig:numerical network}
\end{figure}

The example network under consideration is shown in Figure \ref{fig:numerical network}. This is the same network topology as was considered in \cite{DH_pipe_hydraulic_estimation}, but now with added boundary edges and return network. Edges 1-6 are valves and edges 7-28 are pipes.

The load conditions $t=1\dots T$ used in the numerical examples are generated the following way:

\begin{enumerate}
    \item A set of boundary-flows $\mathring{q}(t)$ are sampled uniformly in the range $\left[100,200\right]$.
    \item Supply-network flows $\Bar{q}(t)$ are calculated through \eqref{eq:supply flows}.
    \item The pressure drop in each supply-network edge is calculated using \eqref{eq:darcy weisbach}.
    \item $p_\alpha - p_\beta$ is randomly selected sufficiently large such that \eqref{eq:pressure is sufficient} is feasible for all valves $v$.
    \item The corresponding feasible valve positions $u_v$ are calculated from \eqref{eq:pressure drop over valve}.
\end{enumerate}

Noisy measurements are considered. If $x$ is a value generated by the above procedure, then the value used in the estimation process will be $\hat{x}$, sampled uniformly in the range $\left[ x(1-\epsilon),x(1+\epsilon)\right]$ where $\epsilon$ denotes the noise level.

The deterministic case $\epsilon=0$ is considered first, followed by the noisy case $\epsilon = 0.01$.

\subsection{Deterministic Measurements}\label{sec:results}
First we estimate the parameters using noise-free measurements under four load conditions. Table \ref{tab:parameter values} presents the actual and estimated hydraulic parameters. Note that the pipe pairs 9-20 and 15-26 differ in parameters between supply and return, contrary to the assumptions of this paper. All parameters are estimated accurately except for these pairs. Note however, that $\hat{s}_9+\hat{s}_{20} = s_9 + s_{20}$ and $\hat{s}_{15}+\hat{s}_{26} = s_{15} + s_{26}$. This is logical since this sum is utilized in reducing \eqref{eq: path alpha beta} to \eqref{eq: path alpha beta simplified}, which defines $\Phi$. If as in \cite{WANG201883} and \cite{DH_pipe_hydraulic_estimation} more pressure-measurements are available, individual parameters should be possible to estimate. However for many control applications such as \cite{AGNER2022100067}, only the sum is required.

\begin{table}[h!]
    \centering
    \caption{Parameter values for each pipe and valve in the network. The estimated values $\hat{s}_i$ are rounded to 3 significant digits, and presented both without noise ($\epsilon=0.0$) and with noise $\epsilon=0.01$). Highlighted in bold and indexed with $^*$ are parameters where the supply-and-return edges have different parameter values.}
    \begin{tabular}{c|c|c|c|c}
        Edge & $s_i$ & $\hat{s}_i$, $\epsilon=0$ & $\hat{s}_i$, $\epsilon=0.01$\\ \hline
        1 & 0.1 & 0.1 & 0.1 & \multirow{6}{4em}{\rotatebox[origin=c]{90}{Valve}}\\
        2 & 0.3 & 0.3 & 0.301 \\
        3 & 0.2 & 0.2 & 0.199 \\
        4 & 0.1 & 0.1 & 0.1 \\
        5 & 0.4 & 0.4 & 0.401 \\
        6 & 0.1 & 0.1 & 0.1 \\ \hline
        7 & 0.0071 & 0.007 & 0.006 & \multirow{6}{4em}{\rotatebox[origin=c]{90}{Supply-pipe}}\\
        8 & 0.00028 & 0.0 & 0.0 \\
        9$^*$ & \textbf{0.0767} & 0.068 & 0.069 \\
        10 & 0.54 & 0.54 & 0.538 \\
        11 & 0.57 & 0.57 & 0.575 \\
        12 & 0.031 & 0.031 & 0.031 \\
        13 & 0.39 & 0.39 & 0.4 \\
        14 & 0.7 & 0.7 & 0.693 \\
        15$^*$ & \textbf{2.067} & 1.828 & 1.835 \\
        16 & 0.39 & 0.39 & 0.397 \\
        17 & 0.64 & 0.64 & 0.627 \\ \hline
        18 & 0.0071 & 0.007 & 0.006 & \multirow{6}{4em}{\rotatebox[origin=c]{90}{Return-pipe}}\\
        19 & 0.00028 & 0.0 & 0.0 \\
        20$^*$ & \textbf{0.059} & 0.068 & 0.069 \\
        21 & 0.54 & 0.54 & 0.538 \\
        22 & 0.57 & 0.57 & 0.575 \\
        23 & 0.031 & 0.031 & 0.031 \\
        24 & 0.39 & 0.39 & 0.4 \\
        25 & 0.7 & 0.7 & 0.693 \\
        26$^*$ & \textbf{1.59 }& 1.829 & 1.835 \\
        27 & 0.39 & 0.39 & 0.397 \\
        28 & 0.64 & 0.64 & 0.627 \\
    \end{tabular}
    \label{tab:parameter values}
\end{table}
\begin{figure*}[h]
    \centering
     \begin{subfigure}[b]{.45\columnwidth}
         \centering
         \includegraphics[width=\textwidth]{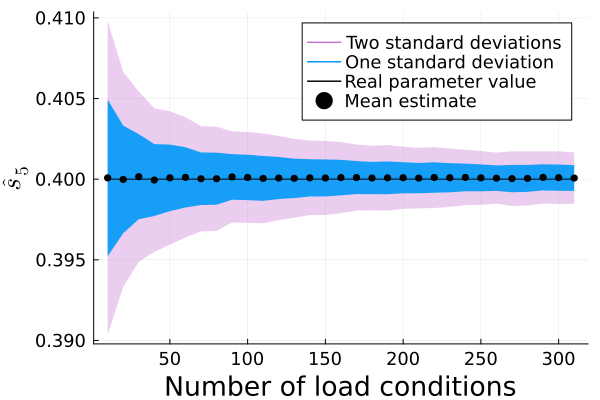}
         \caption{Parameter confidence interval for edge $5$, a boundary edge valve.}
         \label{fig:confidence 5}
     \end{subfigure}
      \hfill
     \begin{subfigure}[b]{.45\columnwidth}
         \centering
         \includegraphics[width=\textwidth]{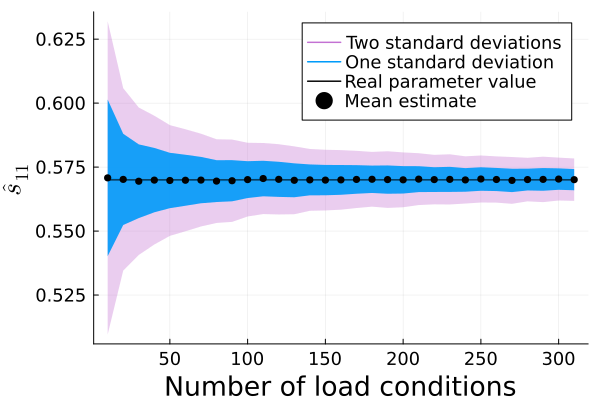}
         \caption{Parameter confidence interval for edge $11$, a supply network pipe.}
         \label{fig:confidence 11}
     \end{subfigure}
     \hfill
     \begin{subfigure}[b]{.45\columnwidth}
         \centering
         \includegraphics[width=\textwidth]{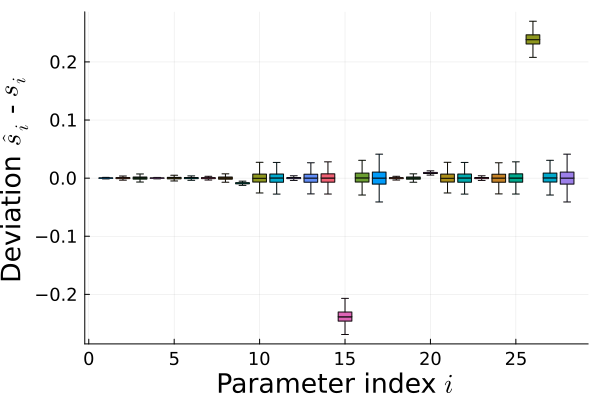}
         \caption{Boxplot of parameter estimates for 50 load conditions.}
         \label{fig:boxplot}
     \end{subfigure}
        \caption{Confidence interval progression for parameters $5$ and $11$ using increasing numbers of load conditions, as well as a boxplot of deviations from the real parameter value using 50 load conditions. Note that edge-pairs 9-20 and 15-26 are not centered, as these edges do not fulfill the assumption that return-and-supply pipes have equal parameter values. The boxplot and each point of the graphs are created using 1000 experiments each.}
        \label{fig:estimation error bars}
\end{figure*}
\subsection{Noisy Measurements}
Table \ref{tab:parameter values} shows also the result of estimation when using $\epsilon=0.01$ (a maximum 1 \% measurement error) and 100 load conditions. The parameters are still determined accurately.

Figures \ref{fig:confidence 5} and \ref{fig:confidence 11} show the confidence interval of the parameter estimates for edges $5$ and $11$ using increasing numbers of load conditions. For each point on the graphs, 1000 estimations are performed. These plots show that both for boundary edges and supply-side edges, the accuracy increases with the number of load conditions. In addition Figure \ref{fig:boxplot} shows the boxplot of estimates using 50 load conditions, showing that all parameters are decided with reasonable accuracy, save for the previously discussed pairs 9-20 and 15-26 where instead the sum is accurate.

\section{Conclusion}\label{sec:conclusion}

In this paper we showed how including a model of valve characteristics can be useful for identifying the hydraulic parameters of a district heating system, such that measurements of all pressures at customer substations are not needed. It was motivated how a set of equations, linear in the hydraulic parameters $s$, can be formed from operational data. A motivation was provided for why the solution $s$ to these equations should be unique in practice. The resulting method was employed on a numerical example, where the estimation procedure was successful even under data gathered within a $1 \%$ measurement error.

\subsection{Future Work}
The function $f$ in this numerical example does not fit all possible component characteristics. For instance, \cite{WANG201883} consider \textit{equal percentage} valves, rather than linear valves. In such a case, one can consider a kernel of functions $f_1,f_2,\dots,f_n$, such that the pressure difference over each edge can be described as
\begin{equation}
\resizebox{.88\hsize}{!}{$p_n-p_m \approx f_1(q_e,u_e)s_{e,1} + f_2(q_e,u_e)s_{e,2} + \dots + f_n(q_e,u_e)s_{e,n}.$}
\end{equation}

More component types can be parameterized and considered. For instance, some boundary nodes could be by pumps, representing other producers in the network. Some of the edges in supply/return-networks could be booster pumps or valves. 

Lastly, a generalization to meshed networks would be highly desirable. This is however difficult, as the linearity of the constraints \eqref{eq:linear constraints short} are based on calculating all edge flows explicitly.

\bibliographystyle{ieeetr}
\bibliography{bibliography.bib}

\begin{thebibliography}{10}

\bibitem{thebible}
S.~Frederiksen and S.~Werner, {\em District heating and cooling}.
\newblock Studentlitteratur, 2013.

\bibitem{lund2018status4th}
H.~Lund, P.~A. Østergaard, M.~Chang, S.~Werner, S.~Svendsen, P.~Sorknæs,
  J.~E. Thorsen, F.~Hvelplund, B.~O.~G. Mortensen, B.~V. Mathiesen, C.~Bojesen,
  N.~Duic, X.~Zhang, and B.~Möller, ``The status of 4th generation district
  heating: Research and results,'' {\em Energy}, vol.~164, pp.~147--159, 2018.

\bibitem{vandermeulen_controlling_2018}
A.~Vandermeulen, B.~van~der Heijde, and L.~Helsen, ``Controlling district
  heating and cooling networks to unlock flexibility: {A} review,'' {\em
  Energy}, vol.~151, pp.~103--115, May 2018.

\bibitem{brange_decision-making_2019}
L.~Brange, K.~Sernhed, and M.~Thern, ``Decision-making process for addressing
  bottleneck problems in district heating networks,'' {\em International
  Journal of Sustainable Energy Planning and Management}, vol.~20, pp.~37--50,
  2019.

\bibitem{hydraulic_optim_meshed}
Y.~Wang, S.~You, H.~Zhang, W.~Zheng, X.~Zheng, and Q.~Miao, ``Hydraulic
  performance optimization of meshed district heating network with multiple
  heat sources,'' {\em Energy}, vol.~126, pp.~603--621, 03 2017.

\bibitem{AGNER2022100067}
F.~Agner, P.~Kergus, R.~Pates, and A.~Rantzer, ``Combating district heating
  bottlenecks using load control,'' {\em Smart Energy}, vol.~6, p.~100067,
  2022.

\bibitem{quovadisWDS}
D.~A. Savic, Z.~S. Kapelan, and P.~M. Jonkergouw, ``Quo vadis water
  distribution model calibration?,'' {\em Urban Water Journal}, vol.~6, no.~1,
  pp.~3--22, 2009.

\bibitem{NetworkParameterEstimationForDHSystem}
G.~Yin, B.~Wang, T.~Sheng, H.~Sun, Q.~Guo, and Z.~Qiao, ``Network parameter
  estimation for district heating system,'' in {\em 2018 2nd IEEE Conference on
  Energy Internet and Energy System Integration (EI2)}, pp.~1--6, 2018.

\bibitem{WANG201883}
N.~Wang, S.~You, Y.~Wang, H.~Zhang, Q.~Miao, X.~Zheng, and L.~Mi, ``Hydraulic
  resistance identification and optimal pressure control of district heating
  network,'' {\em Energy and Buildings}, vol.~170, pp.~83--94, 2018.

\bibitem{DH_pipe_hydraulic_estimation}
Y.~Liu, P.~Wang, and P.~Luo, ``Pipe hydraulic resistances identification of
  district heating networks based on matrix analysis,'' {\em Energies},
  vol.~13, no.~11, 2020.

\end{thebibliography}
                                                   







\end{document}